\documentclass{phyeauth}
\usepackage{graphicx}
\usepackage{amsmath}
\usepackage{amssymb}

\newcommand{\fig}[1]{Fig.~\ref{#1}}

\begin{document}

\begin{frontmatter}

\title{Fano resonances and decoherence in transport through quantum dots}

\author[tu]{Stefan Rotter\thanksref{thank1}},
\author[mar]{Ulrich Kuhl},
\author[tu]{Florian Libisch},
\author[tu]{Joachim Burgd\"orfer}, and
\author[mar]{Hans-J\"urgen St\"ockmann}

\address[tu]{Institute for Theoretical Physics, Vienna University of
Technology, A-1040 Vienna, Austria}
\address[mar]{Fachbereich Physik, Philipps-Univerit\"at Marburg, D-35032
  Marburg, Germany}

\thanks[thank1]{
Corresponding author.\\
E-mail: rotter@concord.itp.tuwien.ac.at}

\begin{abstract}
A tunable microwave scattering device is presented which allows the
controlled variation of Fano line shape parameters in transmission
through quantum billiards. We observe a non-monotonic evolution of resonance
parameters that is explained in terms of interacting resonances.
The dissipation of radiation in the cavity walls leads to decoherence and thus
to a modification of the Fano profile. We show that the imaginary part of the
complex Fano $q$-parameter allows to determine the absorption constant of the
cavity. Our theoretical results demonstrate further that the two decohering
mechanisms, dephasing and dissipation, are equivalent in terms of their effect
on the evolution of Fano resonance lineshapes.

\end{abstract}

\end{frontmatter}


\section{Introduction}

Asymmetric Fano line shapes are an ubiquitous feature of resonance
scattering when (at least) two different pathways connecting the
entrance with the exit channel exist. Fano resonances have been
discussed in many different fields of physics starting
with photoabsorption in atoms \cite{beut,fano35,fano61}, electron
and neutron scattering \cite{adair,simpa63}, Raman
scattering \cite{cer73}, photoabsorption in quantum well
structures \cite{feist97}, scanning tunnel microscopy \cite{mad98},
and ballistic transport through quantum dots
(``artificial atoms'') \cite{goeres00,noeckel94,rotter2,koba,rotter3}
and molecules \cite{guevara,levstein}.
Interest in
observing and analyzing Fano profiles is driven by their high
sensitivity to the details of the scattering process.
For example, since Fano parameters reveal the presence and the nature
of different (non) resonant pathways, they can be used to
determine the degree of coherence in the scattering device.  This is
due to the fact that decoherence may convert Fano resonances
into the more familiar limiting case of a
Breit-Wigner resonance. Furthermore, Fano profiles provide detailed
information on the interaction between nearby resonances leading
to ``avoided crossings'' in the complex plane \cite{heiss91,burg95},
and to stabilization of discrete states in the continuum
(``resonance trapping'' \cite{ro,rott}).
Possible technological applications of Fano resonances
have recently been suggested in Ref.~\cite{bird03}, exploiting 
the transmission resonances
in transport through open quantum dot systems as a means to generate
spin polarization of transmitted carriers.
\begin{figure}[!tbh]
\centering
\includegraphics[draft=false,keepaspectratio=true,clip,width=0.47\textwidth]{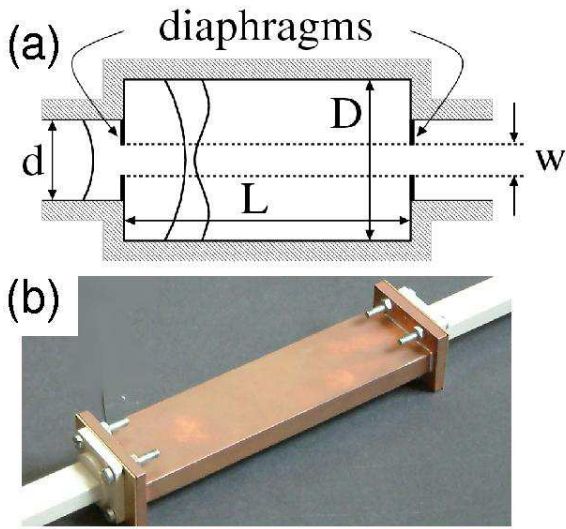}%
\caption{(a) Schematic sketch of the rectangular cavity with leads
attached symmetrically on opposite sides. Exchangeable diaphragms
at the lead junctions allow to control the coupling between
the cavity and the leads. The open even transverse states are
indicated. (b) Photograph of the experimental setup.}
\label{fig:1}
\end{figure}\\Using
the equivalence between the scalar Helmholtz equation for
electromagnetic radiation in cavities with conducting walls and
the Schr\"odinger equation subject to hard-wall boundary
conditions  \cite{stock}, we have designed a scattering device
(Fig.~\ref{fig:1}) that allows the controlled tuning of Fano resonances for
transport through quantum billiards. The evolution of the Fano
parameters as a function of the tuning parameter, in the present
case the degree of opening of the leads, can be traced in
unprecedented detail, since decoherence due to dissipation can be controlled. 
By comparison with
calculations employing the modular recursive Green's function method
(MRGM) \cite{rotter2,rotter}, the parametric variation of Fano resonances and
the degree of decoherence can be quantitatively accounted for.
Furthermore, the relevant pathways can be unambiguously identified in terms of
wave functions representing the contributing scattering channels.
Due to the equivalence between microwave transport and single-electron
motion in two dimensions, our device can be understood as a simulation for
ballistic electron scattering through a quantum dot. In contrast to recent
investigations of mesoscopic dots and single-electron
transistors \cite{goeres00,koba,brems}, where the comparison between theory
and experiment has remained on a mostly qualitative level, our
model system allows for a detailed quantitative analysis of all
features of tunable resonances.

\section{The model}

\begin{figure}[!t]
\centering
\includegraphics[draft=false,keepaspectratio=true,clip,width=.47\textwidth]{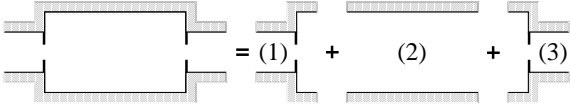}%
\caption{Decomposition of the scattering device into three
 separate substructures: (1) a junction
from a narrow to a wide constriction (with diaphragm), (2) a wide
constriction of length $L$, and (3) a junction from a wide to
a narrow constriction (with diaphragm).}\label{fig:decompose}
\end{figure}

Our microwave scattering device
consists of two commercially available waveguides with
height $h$=7.8\,mm, width $d$=15.8\,mm, and length $l$=200\,mm
which were attached both to the entrance and
the exit side of a rectangular resonator with
height $H$=7.8\,mm, width $D$ = 39\,mm, and length $L$ = 176\,mm, resulting in
a circumference $C$=430\,mm and area $A$=39\,mm$\times$176\,mm
(both in the plane).
At the junctions to the cavity metallic diaphragms of different
openings were inserted (Fig.~\ref{fig:1}). The microwaves with frequencies
between 12.3 and 18.0 GHz, where two even transverse modes are excited
in the cavity and one transverse mode in each of the leads,
are coupled into the waveguide via an adaptor to ensure strong
coupling.\\The experimental results are compared with the predictions
of the MRGM. We solve the $S$ matrix for the single particle Schr\"odinger
equation for this ''quantum dot'' by assuming a constant potential
set equal to zero inside and infinitely high outside of a hard-wall boundary.
At asymptotic distances, scattering boundary conditions are
imposed in the leads.
The coupling of the leads to the cavity of length $L$ can be varied by the
two diaphragms which are placed symmetrically at the two lead openings.
The lead width $d$ and the
width of the rectangular cavity $D$ determine how many flux-carrying
modes are open at a certain energy $\varepsilon$ in each of the
three scattering regions (lead-cavity-lead). We consider in the
following the range of wavenumbers
where one flux-carrying mode is open in each of the leads. Inside the cavity
the first and second even transverse modes are open, thus
providing two alternative pathways of transport through the structure.
In order to characterize the interfering paths, we decompose the
transport across the cavity into a multiple scattering series
involving three pieces \cite{wirt03}, each of which
is characterized by a  mode-to-mode
transmission (reflection) amplitude or a propagator
(see Fig.~\ref{fig:decompose}):
(1) the transmission of the incoming flux from the
 left into the cavity, $t^{(L)}$, or reflection back into the
lead, $r^{(L)}$, (2) the propagation inside the cavity from the left
 to the right, $G^{(LR)}$, or from the right to the left,
 $G^{(RL)}$, and (3) the transmission from
 the interior of the cavity to the right, $t^{(R)}$,
 or internal reflection at each of the two vertical cavity
walls with amplitude $r^{(R)}$. For the Green's functions (i.e.~propagators)
$G^{(LR)}( x_R, x_L )$ and $G^{(RL)}(x_L,x_R)$
we choose a mixed representation which is local in $x$, and
 employs a spectral sum over transverse modes,
 $G^{(LR)}( x_R, x_L )=G^{(RL)}( x_L, x_R )=
 \sum_n\vert n \rangle
 \exp(ik_n |x_R-x_L|)\langle n \vert$,
where $x_{R,L}$ are the $x$-coordinates of the right (left) lead junction
with $|x_R - x_L|= L$. The longitudinal momentum for each
 channel $n$ in the cavity is given by $k_n=\sqrt{k^2-(k_n^c)^2}$,
with the momentum $k=\sqrt{2\varepsilon}$
and the threshold $k$-values $k_n^c=n\pi/D$. Decoherence due to
dissipation of the microwave power in the cavity walls can be
easily incorporated by analytically continuing $k_n$ into the complex plane,
$k_n=\sqrt{k^2-(k_n^c)^2}+i\kappa$.
The shape of Fano resonances
can be used to accurately determine
the degree of dissipation present.\\The multiple scattering expansion
of the transmission amplitude $T$ is then given by
\begin{align}
\label{eqn:3}
 T (k) &=t^{(L)} G^{(LR)} \left\{
 \sum\limits_{n=0}^\infty [ r^{(R)}G^{(RL)} r^{(L)} G^{(LR)} ]^n
 \right\}\, t^{(R)}=\nonumber\\&=t^{(L)} G^{(LR)} \left[1-
r^{(R)}\, G^{(RL)} r^{(L)} G^{(LR)}\right]^{-1} t^{(R)}\,.
 \end{align}
 The identification  of the resonant and non-resonant pathways with
help of Eq.\ (\ref{eqn:3}) is straightforward: due to the absence of
inter-channel mixing in the rectangular (i.e.~non-chaotic) cavity,
the non-resonant contribution corresponds to the $n = 0$ term of
the sum describing direct transmission while the resonant contribution
is made up by all multiple-bounce contributions $(n \geq 1 )$.
The various amplitudes entering Eq.\ (\ref{eqn:3})
can be parametrized in terms of four phases and two moduli \cite{kim01}:
the modulus, $s$, of the reflection amplitude of the wave incoming
in mode 1 and reflected into mode 1 at the left diaphragm,
 \begin{equation}
 \label{eqn:4}
 r_{11}^{(L)} = s e^{i \phi_r}\,,
 \end{equation}
and the modulus, $p$, of the partial injection amplitude of the
incoming wave into the lowest mode of the cavity, corrected for
the partially reflected flux [Eq.~\ref{eqn:4}],
\begin{equation}
\label{eqn:5}
t_{11}^{(L)} = t_{11}^{(R)} = p \sqrt{1-s^2}\,e^{i \phi_t^{(1)}} \, .
\end{equation}
Because of the symmetry of the scattering device, the injection
(ejection) amplitude at the left (right) side are equal. Accordingly,
 the injection amplitude into the second even mode of the cavity is given by
\begin{equation}
\label{eqn:6}
t_{12}^{(L)} = t_{21}^{(R)} = \sqrt{(1-p^2) (1-s^2)}\,e^{i \phi_t^{(2)}} \, .
\end{equation}
Analogous expressions can be deduced  \cite{kim01}
for the other partial amplitudes entering Eq.\ (\ref{eqn:3}). We
omit a detailed analysis of the phases in
Eqs.\ (\ref{eqn:4}-\ref{eqn:6}) since they do not explicitly enter
our analysis in the following. The key observation in the present
context is that the square module $s^2$ is monotonically decreasing
between the limiting values $s^2=1$ for zero diaphragm opening
($w=0$) and $s^2\approx 0$ for fully open diaphragms ($w=d$).
Inserting Eqs. (\ref{eqn:4}-\ref{eqn:6}) into
Eq.\ (\ref{eqn:3}), a closed yet complicated expression for the
transmission probability $|T (\varepsilon, s)|^2$
as a function of the energy $\varepsilon$
and the opening parameter $s$ can be derived \cite{kim01}.
Close to a given resonance $\varepsilon_i^R$ this expression can
be approximated by the Fano form \cite{fano61,kim01},
\begin{equation}
\label{eqn:7}
|T (\varepsilon, s)|^2  \approx \frac{|\varepsilon - \varepsilon_i^R(s) +
q_i (s) \Gamma_i (s) /2|^2}{\left[ \varepsilon - \varepsilon_i^R (s) \right]^2
+ \left[ \Gamma_i (s)/2 \right]^2}\,,
\end{equation}
where $\varepsilon_i^R (s)$ is the position of the $i$-th resonance,
$\Gamma_i (s)$ its width, and $q_i (s)$ the complex Fano asymmetry parameter,
all of which depend on $s$. Window resonances appear in the limit
$q \rightarrow 0$ while the Breit-Wigner limit is reached for $|q| \gg 1$.
Since Fano resonances can
be identified as resulting from the interference between resonances
related to the eigenmodes in the cavity, the parameter $q$
depends very sensitively on the specific position of
the involved resonance poles \cite{lee99,magunov03}.

\section{Comparison between experiment and theory}

\begin{figure}[!tbh]
\centering
\includegraphics[draft=false,keepaspectratio=true,clip,width=.47\textwidth]{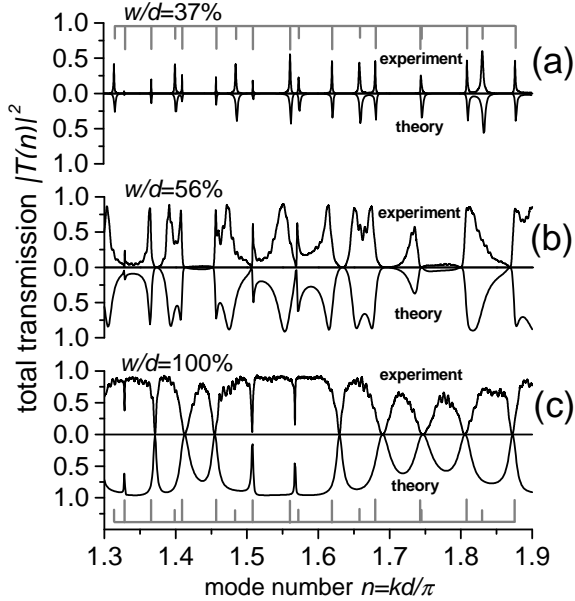}%
\caption{Total transmission probability,
$|T(\sqrt{2\varepsilon}\,d/\pi,w/d)|^2$, for
transport through the rectangular cavity with three different
openings of the diaphragms:
(a) $w/d=37\%$, (b) $w/d=56\%$ and (c) $w/d=100\%$.
For better comparison, the experimental and the calculated results are shown
as mirror images.
The positions of all eigenstates
in the closed cavity are indicated by the gray tick marks.
For all the calculated curves shown, a damping constant of
$\kappa=10^{-4}$~mm$^{-1}$ was used.}
\label{fig:2}
\end{figure}
\begin{figure}[!t]
\centering
\includegraphics[draft=false,keepaspectratio=true,clip,width=.4\textwidth]{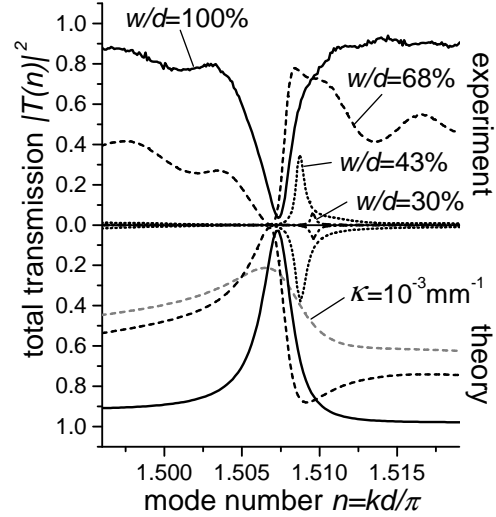}%
\caption{Fano resonance near the
second even excited transverse mode at $kd/\pi\approx 1.5095$.
Experimental and theoretical result for four different cavity
openings ($w/d$) are shown. Curves with equal $w/d$-ratio are displayed
in the same line style (solid, dashed, dotted, dash-dotted). For all
calculated curves a damping factor
$\kappa=10^{-4}$ mm$^{-1}$ was used, except for the additional gray dashed
curve shown for which  $\kappa=10^{-3}$  mm$^{-1}$ and $w/d=0.68$.}
\label{fig:3}
\end{figure}

In Fig.~\ref{fig:2} we present both the
experimental and theoretical dependence of the transmission
probability $|T|^2$ on $k $ (or $\varepsilon$). In the measurement,
the diaphragms were successively closed in steps
of 1\,mm. The data sets of Fig.~\ref{fig:2} (a,b,c) represent the transmission
probability for three different  values of the opening of the diaphragms
$w=5.8,\,8.8,\,$ and 15.8\,mm, respectively. Note the remarkable degree
of agreement between the measured and the calculated data
with only the absorption constant $\kappa$ as an adjustable parameter.
In Fig.~\ref{fig:2}a where $w/d \approx 0.37 $,
transport is suppressed and mediated only by resonance
scattering with narrow
Breit-Wigner shapes centered at the eigenenergies of the closed
billiard as indicated by the tick marks. With increasing diaphragm
opening (Fig.~\ref{fig:2}b) transport acquires a significant
non-resonant contribution,
leading to the widening and the overlap of resonances. Finally,
for fully open leads (Fig.~\ref{fig:2}c), $w/d=1$ (or $s \approx 0$)
resonances appear as narrow window resonances in a non-resonant
continuum. The trajectory of the resonance parameter as a
function of $s$ can be both experimentally and theoretically
mapped out in considerable detail. Different types of resonances
can be identified by their characteristically different resonance
parameters. The evolution of the Fano parameter as a function of
$w/d$ (or $s$) for one resonance is highlighted in Fig.~\ref{fig:3}.
The transition from a narrow Breit-Wigner resonance via a
somewhat wider asymmetric Fano profile to a window resonance
is clearly observable.\\The choice $\kappa=10^{-4}$ mm$^{-1}$ provides us a
good fit and an upper bound of the damping present. Note that
even a slightly larger value of $\kappa=10^{-3}$ mm$^{-1}$
would drastically deteriorate the agreement
between experiment and theory (see Fig.~\ref{fig:3}).
In line with the value $\kappa\approx 10^{-4}$\,mm$^{-1}$,
we obtain an imaginary part of the complex Fano parameter for systems without
time-reversal symmetry \cite{brems} out of our fitting procedure as
$\rm{Im}\, q \lesssim 0.1$.
We note that by using superconducting cavities $\kappa$ could still
be further reduced \cite{rich}, however with
little influence on the result, since we have already
nearly reached the fully coherent
limit.

\section{Evolution of Fano parameters}
\begin{figure}[!b]
\centering
\includegraphics[draft=false,keepaspectratio=true,clip,width=.47\textwidth]{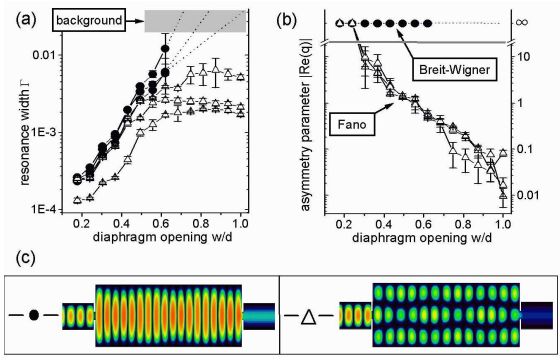}%
\caption{(a) Resonance width $\Gamma$ and (b) real part of the
asymmetry parameter $|{\rm Re}(q)|$ as a function of the diaphragm
opening $w/d$. The data are obtained by fits to experimental results.
Solid circles ${\bullet}$ (empty triangles $\triangle$)
correspond to resonances originating
from the first (second) even cavity eigenstate. Typical wavefunctions
$|\psi(x,y)|^2$ for these two classes of resonances are shown in (c).
In (a) the width $\Gamma$ of the ${\bullet}$-resonances
is monotonically
growing until resonances disappear in the background
of the measured spectra (see gray horizontal bar). For
$\triangle$-resonances $\Gamma$ reaches a local maximum and slightly
decreases for $w/d\rightarrow 1$. In (b) the $\bullet$-resonances
always have a $|{\rm Re}(q)|>10$, above
which the Fano resonances are very
close to the Breit-Wigner lineshape $[{\rm Re}(q)=\infty]$.
For the $\triangle$-resonances
$q$ shows a strong $w/d$-dependence: resonances undergo a complete
evolution from Breit-Wigner to window type as $w/d$ varies
between 0 and 1.}
\label{fig:4}
\end{figure}

Following the parametric evolution of a large number of resonances
yields a characteristic pattern of Fano resonance parameters
(Fig.~\ref{fig:4}a,b).
Obviously, two distinct subsets of resonances appear in the
rectangular billiards: one set is characterized by a strictly monotonic
increase of $\Gamma$ with increasing opening of the cavity and
a large and only weakly dependent asymmetry parameter $q$.
A second set of resonances features a strongly varying $q$
(on the log-scale!) from large values near the Breit-Wigner
limit to values
close to $q \approx 0$ for wide opening, yielding a window
resonance. At the same time, the width $\Gamma$ first increases
with $w/d$ increasing from close to 0, then reaches a
local maximum and finally decreases slightly
when $w/d\rightarrow 1$. A similar non-monotonic behavior
of $\Gamma$ was recently observed in a
single-electron transistor experiment \cite{goeres00}. Such
features can be understood in terms of avoided crossings
in the complex plane \cite{heiss91,burg95} between interacting
resonances. While the von Neumann-Wigner theorem \cite{neu29}
for bound states predicts avoided crossings between states of
the same symmetry and thus a non-monotonic variation of the
eigenenergy, interacting resonances can also display avoided
crossings on the imaginary axis \cite{heiss91,burg95}, i.e.~exchange
of the width of resonances and thus leading to a non-monotonic behavior
of one of the $\Gamma$ involved. The two resonance poles approach
each other in the complex energy plane and undergo an avoided
crossing as a function of the coupling parameter $s$. As a result,
for increasing $s$ the width of the resonance with large $\Gamma$
gets even larger and will
form a background, on top of which the second narrowed resonance is situated.
This somewhat counterintuitive stabilization and
narrowing of the width of one resonance
despite an increased opening, i.e.~an increased coupling to the
environment, is sometimes referred to as ``resonance
trapping'' \cite{ro,rott} and in the limit of $\Gamma \approx 0$
as the ``formation of bound states in the continuum''. This mechnism
could possibly provide an alternative explanation for the results of
Ref.~\cite{goeres00}, where such a non-monotonic behavior was
observed and has been previously attributed to
increased impurity scattering \cite{brems}.\\A very
interesting feature of interacting resonances is the fact
that they can be directly related to scattering wavefunctions
(see Fig.~\ref{fig:4}c).
Resonances that undergo a complete evolution from Breit-Wigner
resonances to a window resonance are all associated with the
{\it second} even excited state in the cavity, while resonances that
display a strictly monotonic increase of the width with increasing
cavity opening
are connected to transport through the transverse {\it ground state}
of the cavity. This mapping is controlled by the amplitude $p$
for transmission through the first transverse mode
[see Eqs.~(\ref{eqn:5},\ref{eqn:6})]. In the case that $p^2>1/2$
all resonances associated with the first mode are broader than
the resonances associated with the excited state and
vice versa for $p^2<1/2$. For geometric reasons
the scattering device studied here (Fig.~\ref{fig:1})
always favors transport through the first cavity mode and therefore
$p^2>1/2$. We thus arrive at the remarkably simple result
that all resonances associated with a first mode feature a strictly
monotonic $\Gamma$ and a large, but weakly varying $q$, while all
resonances associated with the second mode feature a smaller
and non-monotonic $\Gamma$ with $q$ undergoing the complete
evolution from the Breit-Wigner to the window limit.
This one-to-one mapping is also indicated in Fig.~\ref{fig:2},
where only second-mode
resonances (indicated by the dotted tick marks) ``survive'' the transition
of $w/d\rightarrow 1$ while all first-mode resonances (short tick marks)
vanish in the background of the transmission spectrum. This knowledge
could be very useful for the investigation of
electron dynamics in mesoscopic scattering systems where
the parametric evolution of Fano resonances could yield
information about the interaction of internal states and their
coupling to the environment.

\section{Resonance poles in the complex plane}

As discussed above, the transmission amplitudes for scattering through
the rectangular cavity can be parametrized in terms of six independent
parameters \cite{kim01} (provided that no dissipation is present).
By evaluating the resulting expression for the transmission probability
$|T(k_F)|^2$ at complex values of
the Fermi wavenumber $k_F$ we can explicitly
investigate the resonance poles in the complex plane
and their dynamics as a function of the diaphragm opening.
 \begin{figure}[!b]
\centering
\includegraphics[draft=false,keepaspectratio=true,clip,width=.47\textwidth]{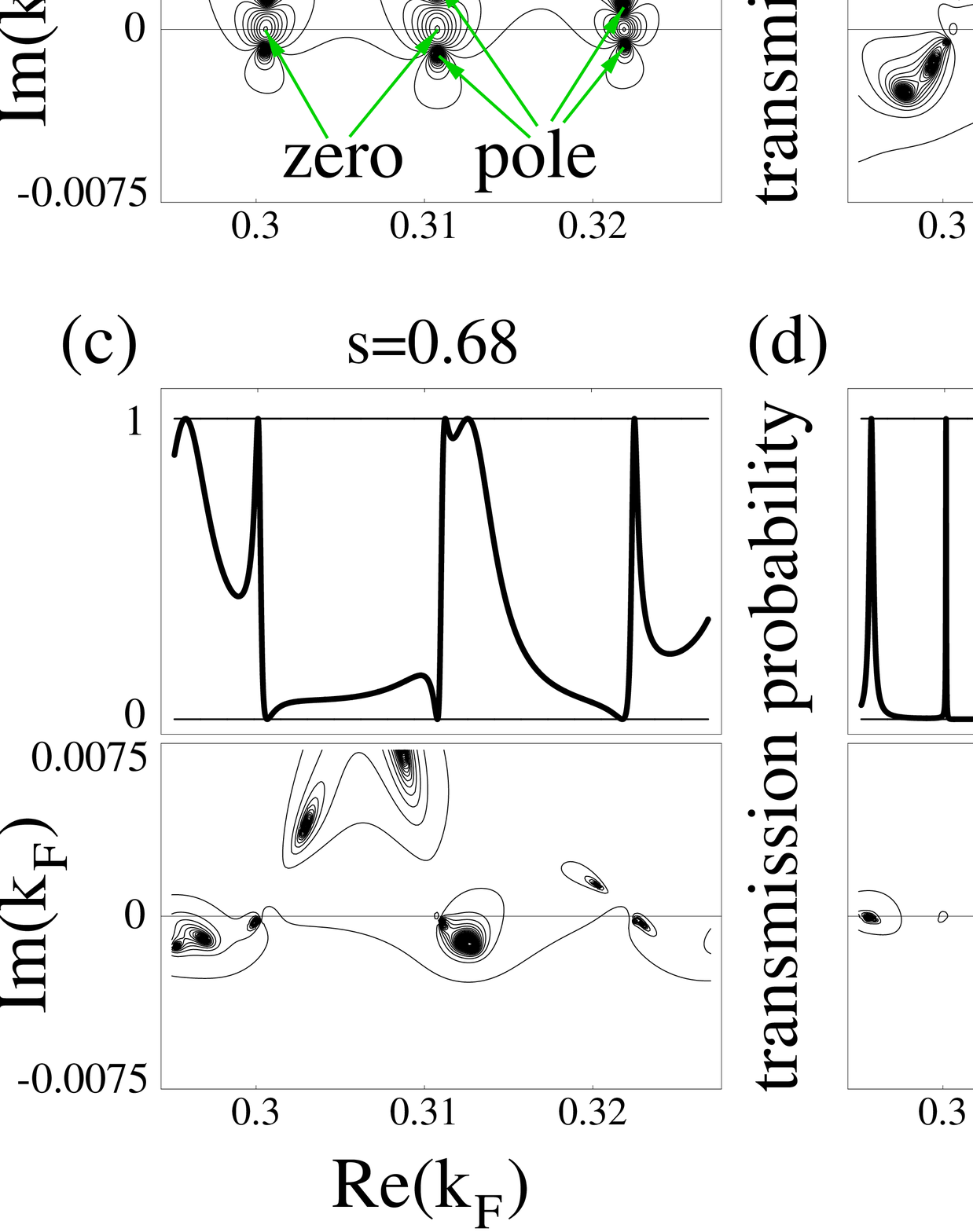}%
\caption{The calculated $k_F$-dependence of transmission
through the rectangular cavity (without dissipation). The diaphragms
are successively closed: (a) $s=0$, (b) $s=0.29$, (c) $s=0.68$,
and (d) $s=0.97$. The upper half of
each figure shows the transmission probability $|T(k_F)|^2$
on the real $k_F$-axis.
The lower half depicts contour plots of the transmission
probability $|T(k_F)|^2$ in the complex $k_F$-plane. Poles and zeros in
transmission move in the $k_F$-plane as a function of the opening of the
diaphragms. Two of the zeros and four of the poles in (a)
are marked by arrows.}\label{fig:5}
\end{figure}
Inspired by the work in \cite{porod93,shao94} we
calculated contour plots of the transmission probability $|T(k_F)|^2$
in the complex $k_F$-plane,
displayed in \fig{fig:5}. In general we can associate with every resonance in
the transmission probability $|T(k_F)|^2$
a pair of poles for $|T(k_F)|^2$ in the complex $k_F$-plane
and a point on the real $k_F$-axis where $|T(k_F)|^2$ goes to zero.
For full opening of the
diaphragms ($s\approx 0$) the complex poles and the points where the
transmission probability goes to zero 
occur at the same Re$(k)$, thus giving rise
to almost symmetric resonances (see \fig{fig:5}a).
When gradually closing the diaphragms, the poles are situated at
a Re$(k)$ which is different from the position of the ``zero''
on the real axis.
Such a configuration gives rise to asymmetric Fano resonances
(see \fig{fig:5}b,c). If the diaphragms are almost shut
($s\approx 1$), we are left with poles situated
very close to the real $k$-axis (see \fig{fig:5}d). In line with this
observation, the resulting
resonances are very narrow and isolated from each other.

\section{Decoherence: Dephasing versus dissipation}

Decoherence in an open quantum system can be caused by different mechanisms:
dephasing and dissipation. Both of them are contained in the relaxation
operator in a quantum Liouville equation. Decoherence in the microwave
scattering device is caused by dissipation, i.e.~the absorption of microwave
energy by the walls. As discussed above, the absorption constant can be
extracted from the shape of the Fano profile in terms of a complex $q$
parameter. 
The good agreement with theory allows
to accurately determine the degree of damping and thus of decoherence 
present in the experiment. As the Fano profile, in particular near its
minimum, is very sensitive to any non-interfering incoherent
background, we can determine an upper bound for the damping
by comparison between experiment and theory to be
$\kappa \lesssim 10^{-4}$ mm$^{-1}$. The consistency of this approach can be
checked by an independent calculation of the absorption 
via the skindepth $\delta$.
The cavity in the experiment is made out of oxygen-free copper (DIN 1787:SF-Cu)
with a density $\rho\approx$\,8.9\,kg/dm$^{-3}$ and a
conductivity $\sigma\approx\,4.9\times 10^{4}$ Ohm$^{-1}\,$mm$^{-1}$ at room
temperature. The skin depth is given by \cite{jackson}
\begin{equation}
  \delta = \frac{1}{\sqrt{\sigma \pi \mu \nu}}\,,
\end{equation}
with $\nu$ being the microwave frequency and
$\mu$ the susceptibility of the metal,
$\mu\approx\mu_0=4\pi 10^{-7}$ Vs (Am)$^{-1}$.
The quality $Q$ of a resonator is defined as
the ratio of the eigenfrequency $\nu$
to its width $\Delta\nu$.
The $Q$ can be calculated as \cite{jackson},
\begin{equation}
  Q=\frac{\mu}{\mu_0} \frac{H}{\delta}
\frac{1}{2\times(1+\xi \,\frac{C H}{4A})}\,,
\end{equation}
where the height $H$, the circumference $C$, and the area $A$ of the cavity
enter explicitly. The constant
$\xi$ is a geometrical factor close to unity.
For the frequency range used in the experiment we find a frequency width
$\Delta\nu=\nu/Q\approx\,$2.5 MHz. Note that this width 
originates purely in the absorption of
microwaves in the cavity walls, as opposed to the resonance broadening caused
by the openness of the diaphragms. In line with this result we find that
the above frequency width of 2.5 MHz corresponds very well to 
the resonance width in the case of small diaphragm opening 
(i.e. $w/d\lesssim$30\%), where resonances take on the Breit-Wigner lineshape. 
The Fourier transform of this resonance form
is an exponentially decaying function in coordinate
space, the decay constant of which is given by $\kappa$.
We can therefore directly determine 
$\kappa=\Delta\nu\times\pi/c\,$, where $c$ is the speed of light.
This finally yields the estimate
$\kappa\approx 2.7\times 10^{-5}$ mm$^{-1}$.
The value obtained from the fit of the Fano resonances is somewhat larger
($\kappa\approx 1\times 10^{-4}$ mm$^{-1}$),
which is probably due to additional absorption in leads.
It has been shown in
Ref.~\cite{brems} that Fano resonances acquire an imaginary $q$-parameter
under the influence of dephasing.  Our results presented here
clearly indicate that dissipation has the same
interference-reducing (i.e.~decohering) effect on a Fano resonance as
dephasing does. In order to illustrate the equivalence of the decohering
effect of dephasing and dissipation on Fano resonances we simulate 
dephasing in cavity explicitly
and compare with results on dissipative transport.
Whereas absorption of microwaves is included in our model by adding to
the real wavenumber in longitudinal direction an imaginary part,
$k_n=\sqrt{k^2-(k_n^c)^2}+i\kappa$, the effect of dephasing is taken into
account in the following way: For each of the traversals of the cavity
in the multiple scattering expansion of Eq.~(\ref{eqn:3})
a Gaussian distributed random phase $\phi_{\rm ran}\in{\mathbb R}$
is added to the longitudinal wavenumer,
$k_n=\sqrt{k^2-(k_n^c)^2}+\phi_{\rm ran}$ appearing in the 
constant energy propagators $G^{(LR)}$ and $G^{(RL)}$.
The degree of dephasing is controlled by the width of the Gaussian
distribution. The numerical data for transmission are then
averaged over a large ensemble of different realizations of the random
phase.\\Note that dephasing, as well as dissipation,
causes the unitarity of the scattering process to be violated. 
Dissipative absorption of microwaves in the cavity walls leads to a net loss
of microwave radiation in the system with $U(k)=|T(k)|^2+|R(k)|^2<1$,
dephasing causes $U(k)$ to fluctutate {\it around}
the unit value. For a meaningful
comparison we enforce unitarity simply
by renormalizing the transmission and reflection
probabilities explicitly, $|T(k)|^2\rightarrow |T(k)|^2/U(k)$ and
$|R(k)|^2\rightarrow |R(k)|^2/U(k)$. We note that rigorous methods for
restoring unitarity in open decohering quantum systems are available for both
time-independent \cite{brems} and time-dependent systems \cite{marek}. 
 \begin{figure}[!t]
\centering
\includegraphics[draft=false,keepaspectratio=true,clip,width=.47\textwidth]{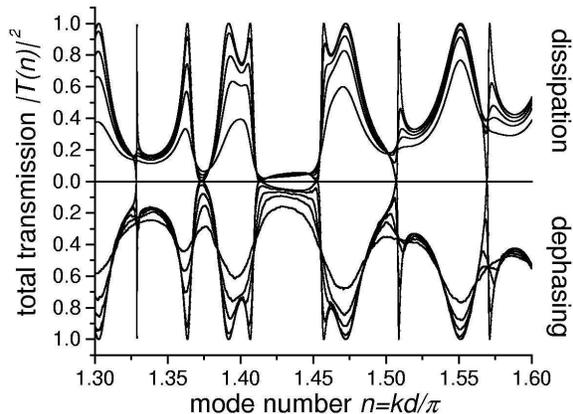}%
\caption{Renormalized transmission probability,
$|T(\sqrt{2\varepsilon}\,d/\pi,
w/d)/U(\sqrt{2\varepsilon}\,d/\pi,w/d)|^2$, for
transport through the rectangular cavity under the influence of dissipative
and dephasing mechanisms, respectively. Opening ratio of the
diaphragms $w/d=56\%$. Above the horizontal axis curves with
different dissipation constants $\kappa$ are shown: $\kappa= 0,5,11,17,28\times
10^{-4}$ mm$^{-1}$. The data below the axis contain
a random phase factor for each longitudinal crossing of the
cavity. The depicted curves feature each a different variance
$\Delta k$ of the Gaussian random phase
distribution: $\Delta k=0,11,20,28,45\times 10^{-4}$ mm$^{-4}$.
Note that both decohering mechanisms
have a very similar effect on the Fano resonances.}\label{fig:6}
\end{figure}
Fig.~\ref{fig:6} illustrates that dephasing and dissipation indeed
have a very similar effect on the evolution of Fano resonances and therefore
yield complex values of  the $q$-parameter. One conclusion we draw from
this finding is that however useful Fano resonances are as a means to analyze
the decoherence present in a scattering process, they do not
allow to distinguish between different mechanisms leading to decoherence.

\section{Summary}
In summary, the rectangular microwave cavity attached to two
leads allows to study the interplay between resonant and
non-resonant transport in unprecedent detail. By a controlled
change of the opening, tuning a Fano resonance from the Breit-Wigner
limit to the window resonance limit has become possible.
Fano resonances can be used to accurately
determine the degree of decoherence present in a scattering
device. We show that dephasing and dissipative mechanisms have very
similar effects on the evolution of Fano resonances.
The non-monotonic behavior of resonance parameters can
be related to avoided crossings between interacting resonances,
which can be unambiguously associated with different
resonant modes of the cavity. The latter feature is a
consequence of the separability of the wave function in the
closed cavity. Future investigations along these lines for non-separable
chaotic cavities promise new insights into the resonance dynamics
of open chaotic systems.\\
\linebreak
{\bf Acknowledgements\\}
\linebreak
We thank D.-H.~Kim, K.~Kobayashi,
C.~M\"uller, E.~Persson, I.~Rotter, C.~Stampfer,
and L.~Wirtz for helpful discussions. Support by the Austrian Science
Foundation (FWF-SFB016) is gratefully acknowledged.


\end{document}